\documentstyle[12pt]{article}
\def\ie{{\em i.e.}}
\def\eg{{\em e.g.}}

\def\beq{\begin{equation}}
\def\eeq{\end{equation}}

\catcode`\@=11 

\def\lsim{\mathrel{\mathpalette\@versim<}}
\def\gsim{\mathrel{\mathpalette\@versim>}}
\def\@versim#1#2{\vcenter{\offinterlineskip
    \ialign{$\m@th#1\hfil##\hfil$\crcr#2\crcr\sim\crcr } }}
\def\etal{{\em et. al.}}
\def\JL{J. L. Lopez}
\def\DVN{D. V. Nanopoulos}

\def\t1{{\tilde 1}}

\def\GeV{\,{\rm GeV}}

\def\to{\rightarrow}
\def\pb{\,{\rm pb}}
\def\ipb{\,{\rm pb}^{-1}}

\def\NPB#1#2#3{Nucl. Phys. B {\bf#1} (19#2) #3}
\def\PLB#1#2#3{Phys. Lett. B {\bf#1} (19#2) #3}

\def\PRD#1#2#3{Phys. Rev. D {\bf#1} (19#2) #3}
\def\PRL#1#2#3{Phys. Rev. Lett. {\bf#1} (19#2) #3}

\def\hepph#1{{\tt hep-ph/#1}}
\def\hepth#1{{\tt hep-th/#1}}

\begin{document}
\begin{flushright}
\baselineskip=12pt
CTP-TAMU-01/96\\
DOE/ER/40717--24\\
ACT-01/96\\
\tt hep-ph/9601261
\end{flushright}

\begin{center}
\vglue 1cm
{\Large\bf Experimental constraints on a stringy SU(5)$\times$U(1) model\\}
\vglue 0.75cm
{\Large Jorge L. Lopez$^1$, D.V. Nanopoulos$^{2,3,5}$, and A. Zichichi$^{4,5}$\\}
\vglue 0.5cm
\begin{flushleft}
$^1$Department of Physics, Bonner Nuclear Lab, Rice University\\ 6100 Main
Street, Houston, TX 77005, USA\\
$^2$Center for Theoretical Physics, Department of Physics, Texas A\&M
University\\ College Station, TX 77843--4242, USA\\
$^3$Astroparticle Physics Group, Houston Advanced Research Center (HARC)\\
The Mitchell Campus, The Woodlands, TX 77381, USA\\
$^4$CERN, 1211 Geneva 23, Switzerland\\
$^5$University and INFN, Bologna, Italy
\end{flushleft}
\end{center}

\vglue 0.75cm
\begin{abstract}
We re-examine the phenomenological aspects of a recently proposed string-derived SU(5)$\times$U(1) one-parameter supergravity model, and explore the sensitivity of the model predictions to variations in the strong coupling and the string unification scale. We also perform an analysis of the constraints on the parameter space of the model in light of the recent Tevatron  trilepton data and the LEP~1.5 chargino and slepton data. We obtain $m_{\chi^\pm_1}\gsim70\GeV$, which excludes one-third of the parameter space of the model. The remainder of the parameter space should be probed by
ongoing analyses of the Tevatron trilepton data and forthcoming LEP~2 runs.
\end{abstract}
\vspace{1cm}
\begin{flushleft}
\baselineskip=12pt
CTP-TAMU-01/96\\
DOE/ER/40717--24\\
ACT-01/96\\
January 1996\\
\end{flushleft}
\newpage
\setcounter{page}{1}
\pagestyle{plain}
\baselineskip=14pt

\section{Introduction}
\label{sec:introduction}
String model-building has come a long way since its inception about a decade ago, making the calculation of experimental predictions in specific string models an ever more refined art. For instance, in free-fermionic string models,
after specifying the requisite two-dimensional world-sheet inputs that define
a model (\ie, basis vectors and GSO projections), it was originally only
known how to obtain the spectrum of states and the corresponding gauge group.
Later one learned how to calculate the superpotential interactions at cubic
and non-renormalizable level. More recently one has been able to identify
the fields that parametrize flat directions of the scalar potential (moduli) in this class of models, a necessary step in the calculation of the K\"ahler potential. All of this stringy ``technology" has been applied to search for
models with vanishing vacuum energy at the tree- and one-loop levels \cite{LN}.
In Ref.~\cite{Zero} we presented a string-derived model based on the gauge
group SU(5)$\times$U(1) which incorporated these theoretical advances. This
model predicted the top-quark mass to be $m_t\approx175\GeV$, and a rather light spectrum of superparticle masses expressed as a function of a single mass parameter. Here we discuss recent theoretical and experimental developments
that have increased the motivation for such class of models, and refine and update the corresponding experimental predictions in light of the recent Tevatron and LEP~1.5 data.

First let us discuss the theoretical developments. The model in Ref.~\cite{Zero} possessed vanishing vacuum energy (at tree-level) before and after spontaneous supersymmetry breaking. It has since been shown that this essential property survives the shift in vacuum expectation values of scalar fields required to cancel the anomalous $\rm U_A(1)$ contribution to the D-term in realistic string models \cite{Erice95}. Also, the model presented 
in Ref.~\cite{Zero} suffered from a quadratically divergent contribution to the one-loop vacuum energy: ${1\over16\pi^2}\,Q_0\, m^2_{3/2}M^2_{Pl}$, with $Q_0=4$, although it was argued that such value was ``small" enough to be possibly shifted towards zero by high-order effects. It has now been shown that in the process of cancelling the anomalous $\rm U_A(1)$, $Q_0$ is shifted to $Q=Q_0+\Delta Q$, such that $Q=0$ can be naturally obtained in this model \cite{StrM^2,Erice95}. This shift is only effective when $V_0=0$ and $Q_0$ is sufficiently small. These developments put the model in Ref.~\cite{Zero} (the only known one with $V_0=Q=0$) on much firmer theoretical ground, making exploration of its consequences a better motivated and pressing endeavor.

On the experimental side, the prediction for the top-quark mass in this model 
($m_t\approx175\GeV$) agrees rather well with experimental observations. Also, the predicted light charginos and neutralinos ($m_{\chi^\pm_1}\approx m_{\chi^0_2}\lsim90\GeV$) entail dilepton and trilepton signals at the Tevatron that are now entering the range of experimental sensitivity. These could also be probed at LEP~1.5 (the first upgraded phase of LEP with $\sqrt{s}=130-136\GeV$), as could the light right-handed sleptons ($m_{\tilde\ell_R}\lsim50\GeV$). It has also become apparent that the predicted value for $\alpha_s(M_Z)$ in the traditional minimal SU(5) GUT is uncomfortably large ($\alpha_s(M_Z)>0.123$ \cite{baggeretal}), and certainly cannot explain the low-energy determination of $\alpha_s(M_Z)$ (around 0.11). On the other hand, it has been recently shown \cite{Lowering} that SU(5)$\times$U(1) can naturally explain the whole range of $\alpha_s$ values being considered, motivating further studies of models with SU(5)$\times$U(1) gauge symmetry.

In what follows we first study the robustness of the scenario presented in
Ref.~\cite{Zero} under variations in the string unification scale, the strong
gauge coupling [$\alpha_s(M_Z)$], and two-loop effects in the running of the
gauge couplings (Sec.~\ref{sec:refinements}). We then turn to the constraints
on the parameter space of the model that follow from recent Tevatron trilepton
data (Sec.~\ref{sec:Tevatron}) and LEP~1.5 chargino and slepton data (Sec.~\ref{sec:LEP15}). We summarize our conclusions in Sec.~\ref{sec:conclusions}.

\section{Running refinements}
\label{sec:refinements}
In the present scenario of gauge coupling unification the Standard Model gauge couplings are run up to the string scale, where SU(5)$\times$U(1) is assumed to break down to the Standard Model gauge group \cite{LNZI}. This scenario predicts the existence of intermediate-scale vector-like particles:
a pair of $(Q,\bar Q)$ [$({\bf3},{\bf2}),(\bar{\bf3},{\bf2})$] and a pair of $(D,\bar D)$ [$({\bf3},{\bf1}),(\bar{\bf3},{\bf1})$], such that string unification occurs at the string scale. The masses of these particles can be in principle derived from the model \cite{search}, although in practice they are adjusted to fit this scenario. These masses, and the whole superparticle spectrum, depend on the value of $\alpha_s(M_Z)$ used in the calculations. They also depend on the value chosen for the string scale. Previously we simply set $M_{\rm string}=10^{18}\GeV$. The value of the top-quark mass also depends on these two inputs. Here we refine our calculations of these observables by allowing a wide range of $\alpha_s(M_Z)$ values consistent with low-energy determinations ($0.108-0.110$) and the world-average ($0.118\pm0.006$). We also 
study the effects of using the proper string scale $M_{\rm string}=5\times g\times10^{17}\GeV$, where $g$ is the gauge coupling at the unification scale. Our previous choice for $M_{\rm string}$ gives an indication of the effects of string thresholds, which in this class of models tend to yield a slightly larger effective unification scale \cite{Lacaze}. Recent more refined analyses
also entail small shifts in the effective string unification scale \cite{KK}.

\begin{figure}[p]
\vspace{6.5in}
\includegraphics{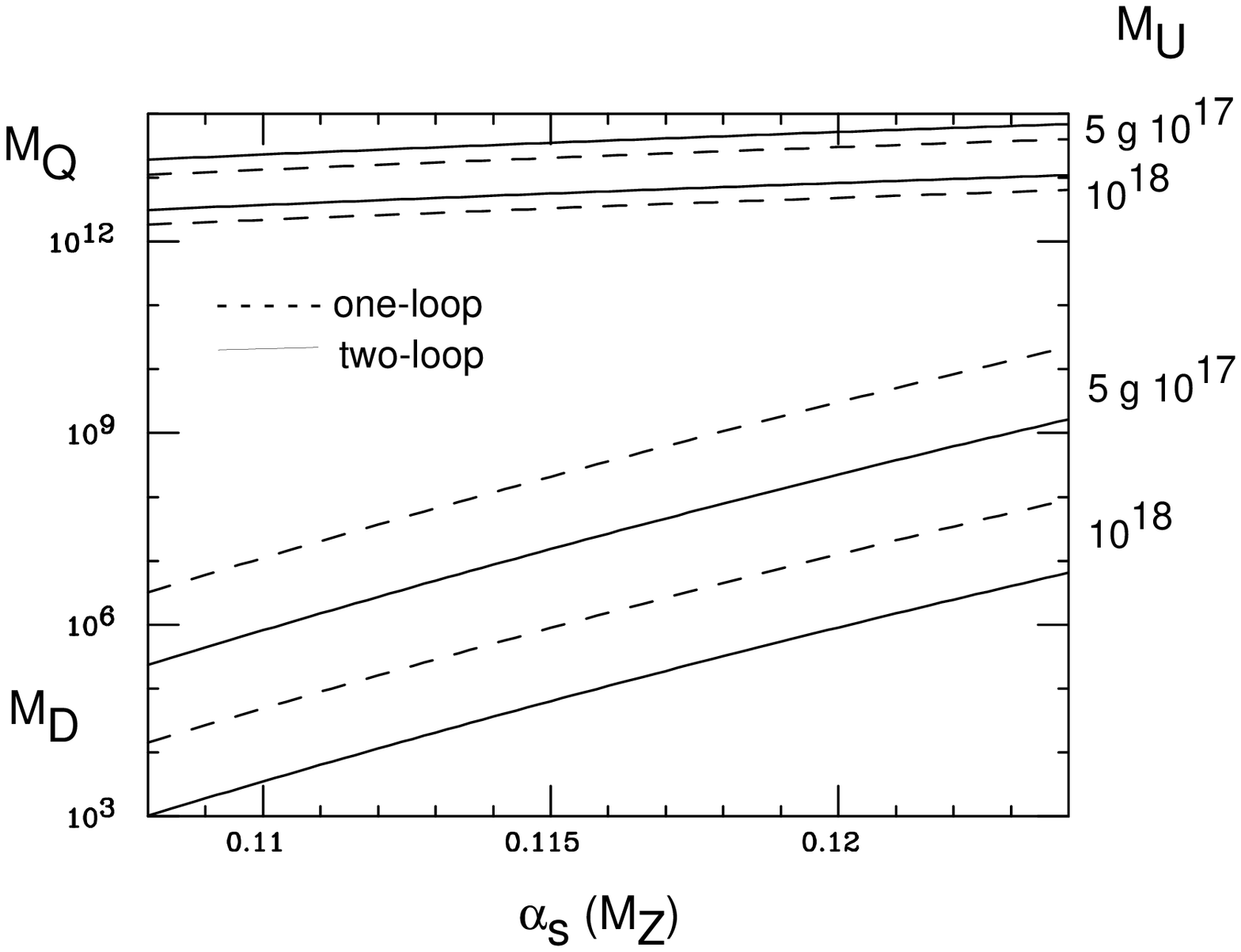}
\caption{The masses of the $Q$ and $D$ intermediate-scale particles as a
function of $\alpha_s(M_Z)$ using one-loop (dashed lines) and two-loop (solid
lines) contributions to the renormalization group equations. Two choices of the string unification scale ($M_U$) are shown.}
\label{fig:masses}
\end{figure}

\begin{figure}[p]
\vspace{6.5in}
\includegraphics{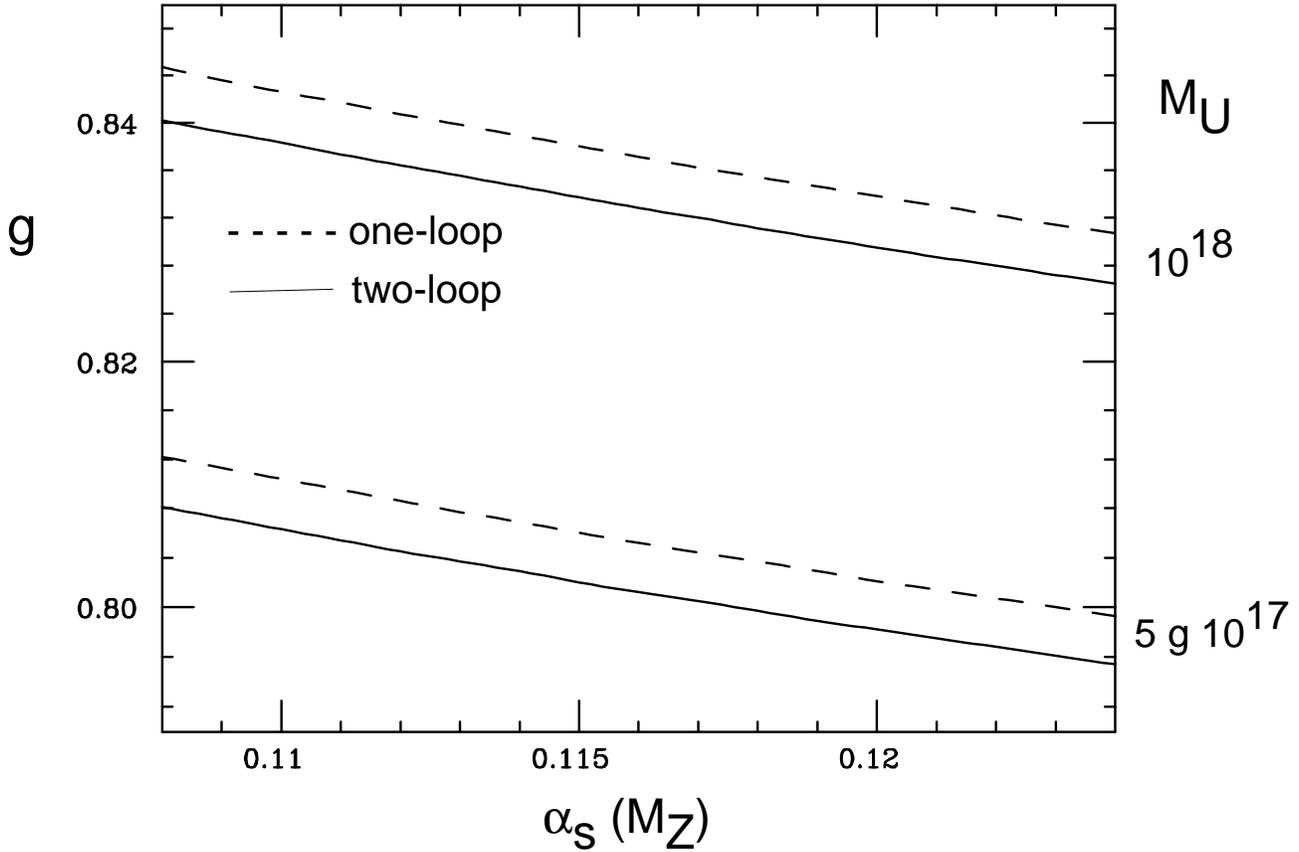}
\caption{The gauge coupling at the string unification scale obtained using one-loop (dashed lines) and two-loop (solid lines) contributions to the renormalization group equations. Two choices of the string unification scale ($M_U$) are shown.}
\label{fig:g}
\end{figure}

\begin{figure}[p]
\vspace{6.5in}
\includegraphics{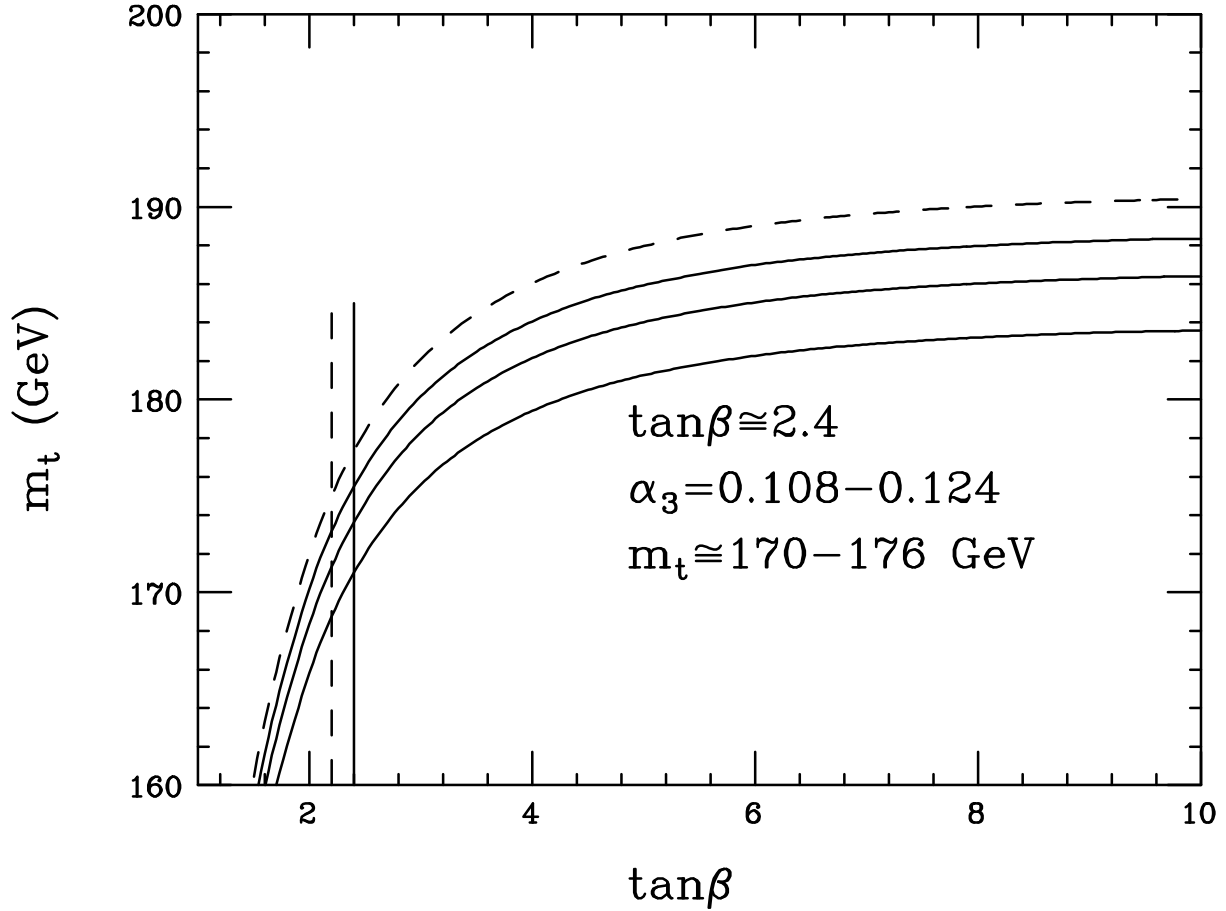}
\caption{The calculated values of the top-quark mass as a function of $\tan\beta$ for $\alpha_s=0.108,0.118,0.124$ (bottom,central,top solid lines)
and $M_U=5\times g\times10^{17}\GeV$. Also shown (for $\alpha_s=0.118$) is the
effect of taking $M_U=10^{18}\GeV$ (dashed line). The vertical lines indicate the dynamically-determined value of $\tan\beta$. Note the stability of the $m_t$ prediction.}
\label{fig:t-zeroU}
\end{figure}

First we study the dependence of $M_Q$, $M_D$, and $g$, as we use the proper string scale, as we take $\alpha_s(M_Z)=0.108-0.124$, and as we allow for two-loop corrections to the running of the gauge couplings. The results for $M_Q,M_D$ and for $g$ are shown in Figs.~\ref{fig:masses} and \ref{fig:g} respectively. In the case of $M_Q$, the largest effect is the shift in the
string unification scale, with small $\alpha_s$ and two-loop dependences. In the case of $M_D$, the largest effect is the dependence on $\alpha_s$, then comes the effect of shifting the string unification scale, and lastly the two-loop effects. The effects on $g$ also follow a hierarchy, although they do not amount to more than a few percent: largest is the effect of shifting the unification scale, then comes the effect of varying $\alpha_s$, and finally the two-loop effects. The shifts in the above quantities impact the calculation of the top-quark mass itself,
\begin{equation}
 m_t=\left({v\over\sqrt{2}}\right)\lambda_t(m_t)
{\tan\beta\over\sqrt{\tan^2\beta+1}}\ ,
\label{eq:mt}
\end{equation}
where $\lambda_t(m_t)$ is obtained from the string-scale prediction of
$\lambda_t(M_U)=g^2$ \cite{Zero}. The above refinements affect $m_t$ through the shifts in $g$, the variations in the running (starting from a different $M_U$, going through the different $Q$ and $D$ thresholds), and the calculated value of $\tan\beta$. The range of (pole) $m_t$ values is shown in Fig.~\ref{fig:t-zeroU} for three values of $\alpha_s$, and $M_U=5\times g\times10^{17}\GeV$. For $\alpha_s=0.118$ we also show the result for $M_U=10^{18}\GeV$. We see that without specifying the value of $\tan\beta$ we obtain 
\begin{equation}
m_t\approx(160-190)\GeV\ ,
\label{eq:mtrange}
\end{equation}
irrespective of the various uncertainties in the calculation. 

In Ref.~\cite{Zero} we showed how the value of $\tan\beta$ can be determined
in terms of the one parameter in the model: one adjusts $\tan\beta$ until
the predicted value of $B_0$ is reproduced by the calculated value of $B_0$
(obtained from the radiative electroweak breaking constraint). This procedure
gives values of $\tan\beta$ which vary with the one model parameter, but only
slightly. For $\alpha_s=0.118$ and $M_U=10^{18}\GeV$ we obtained $\tan\beta\approx2.2-2.3$. The $\tan\beta$ range is cut-off, as are the ranges of all the sparticle masses, by a cut-off in the one parameter in the model.
This cut-off is obtained when the masses of the right-handed charged sleptons 
($\tilde\ell^\pm_R$) become lighter than the mass of the lightest neutralino, entailing a cosmologically unacceptable lightest supersymmetric particle (LSP) \cite{EHNOS}.

Recalculating the values of $\tan\beta$ that are obtained for $\alpha_s=0.108,0.118,0.124$ and $M_U=5\times g\times10^{17}\GeV$ we find that the previous results remain qualitatively unchanged, and quantitatively only slightly modified. The $\tan\beta$ range is now a little higher: $\tan\beta\approx2.35-2.45$, but the top-quark mass as a function of $\tan\beta$ is a little lower (c.f. central solid versus dashed line in Fig.~\ref{fig:t-zeroU}). The result of these two compensating
effects is a rather stable prediction for the top-quark mass in this model
\begin{equation}
m_t\approx(170-176)\GeV\ .
\label{eq:result}
\end{equation}

The spectrum of superparticle and Higgs boson masses is quite close to that
obtained previously in Ref.~\cite{Zero}, and definitely indistinguishable from it given the inherent uncertainties in this type of calculations.  In what
follows we simply use the spectrum obtained in Ref.~\cite{Zero}. In particular
one finds
\begin{eqnarray}
m_{1/2}&\lsim&180\GeV\\
m_{\chi^\pm_1}&\approx& m_{\chi^0_2}\approx(60-90)\GeV\\
m_h&\approx&(80-90)\GeV\\
m_{\tilde\ell_R}&\approx&(45-50)\GeV\\
m_{\tilde q}&\approx&0.98\,m_{\tilde g}
\end{eqnarray}
Note that for specific values of $\alpha_s$, subsets of the
mass intervals above will be realized.

\section{Tevatron constraints}
\label{sec:Tevatron}
\begin{figure}[p]
\vspace{6in}
\includegraphics{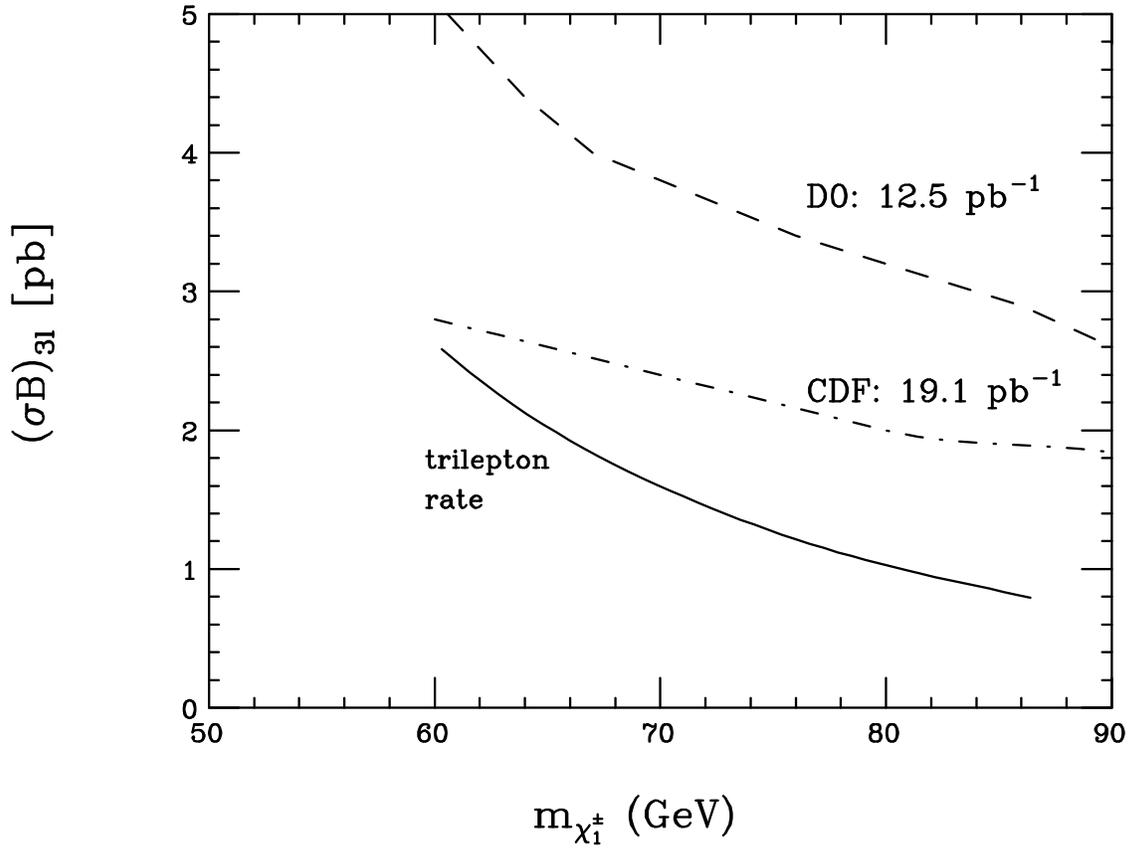}
\caption{Trilepton rates (solid line) at the Tevatron (summed over $eee$, $ee\mu$, $e\mu\mu$, $\mu\mu\mu$) versus the chargino mass originating from chargino-neutralino production. The dashed (dotdashed) line represents the D0 (CDF) upper limit based on $12.5\,(19.1)\,{\rm pb}^{-1}$ of data. The bends
on the experimental curves reflect changes in trigger rates and detection
efficiencies.}
\label{fig:3l}
\end{figure}

The cross section for production of chargino-neutralino pairs and subsequent decay into trileptons ($p\bar p\to\chi^\pm_1\chi^0_2X$; $\chi^\pm_1\to\ell^\pm\nu_\ell\chi^0_1$, $\chi^0_2\to\ell^+\ell^-\chi^0_1$) in
this model has been given in Ref.~\cite{Zero}, and is reproduced in Fig.~\ref{fig:3l}. We recall that the chargino ($\chi^\pm_1$) branching ratio into leptons ($e+\mu$) is $\approx0.5$, whereas that into jets is $\approx0.25$, for all points in parameter space. Also, the neutralino ($\chi^0_2$) decays exclusively to dileptons because of the dominant two-body decay mode $\chi^0_2\to\tilde\ell_R^\pm\ell^\mp$, and therefore the trilepton signal is nearly maximized. The trilepton rates in Fig.~\ref{fig:3l} have been summed over the four channels $eee$, $ee\mu$, $e\mu\mu$, and $\mu\mu\mu$. The D0 Collaboration has released its first official results on these searches based on $12.5\ipb$ of data \cite{D03l}. Their results are conveniently expressed as an upper bound on the trilepton rate (into any one of the four channels) as a function of the chargino mass. The upper limits range from\footnote{These values are the result of multiplying by four the explicit limits given in Ref.~\cite{D03l}, in order to account for our summing over the four possible channels.} $12.4\pb$ for $m_{\chi^\pm_1}=45\GeV$ down to $2.4\pb$ for $m_{\chi^\pm_1}=100\GeV$. The corresponding curve is shown in Fig.~\ref{fig:3l} (dashed line), and does not constrain the model in any way. CDF has also released an upper limit on the trilepton rate based on $19.11\ipb$ of data \cite{Teruki}, that is also shown in Fig.~\ref{fig:3l} (dotdashed line),\footnote{The CDF upper limit shown here is an (adequate) approximation to the actual experimental result, which has some dependence on the decay kinematics through the detection efficiencies.} and which comes quite close
to the model predictions.

Ongoing analyses by CDF and D0 should be able to probe some of the parameter space of the model. Indeed, since few background events are expected, to estimate the reach obtained by examining the full data set ($\sim100\ipb$)
one could simply scale down the present D0 (CDF) upper limit by a factor of $100/12.5\,(19)\approx8\,(5)$. Such sensitivity would appear to be enough to
falsify the model. However, this will likely not be the case, because one of
the leptons from $\chi^0_2$ decay ($\chi^0_2\to\tilde\ell^\pm_R \ell^\mp$,
$\tilde\ell^\pm\to\ell^\pm\chi^0_1$) becomes increasingly softer as the edge
of parameter space ({\em i.e.}, $m_{\tilde\ell_R}=m_{\chi^0_1}$) is approached.
Assuming negligible experimental detection sensitivity for $m_{\tilde\ell_R}-
m_{\chi^0_1}\lsim6\GeV$, only $m_{\chi^\pm_1}\lsim70\GeV$ could be probed. Such
kinematical accidents remind us that ``the absence of
evidence is not evidence of absence." Analogous dilepton searches \cite{di-tri} are not expected to be as sensitive \cite{Zero}, although they will likely boost the supersymmetric signal should trilepton events be observed.

\section{LEP 1.5 constraints}
\label{sec:LEP15}
The intermediate-energy upgrade of LEP (``LEP~1.5") accumulated close to
$3\ipb$ of data (per experiment) at both $\sqrt{s}=130\GeV$ and $136\GeV$.
Preliminary physics results have been recently announced \cite{LEP15}. Two
of the new physics searches are relevant to the present analysis: searches
for chargino pair production and searches for charged-slepton pair production.

Chargino production proceeds via $s$-channel $\gamma,Z$ exchange and $t$-channel sneutrino ($\tilde\nu$) exchange, with the two amplitudes interfering destructively if the sneutrino mass is not too large. In our model we find $m_{\tilde\nu}=(85-115)\GeV$, and the negative interference is not very pronounced, yielding chargino cross sections of a few pb. Experimentally one finds that as long as $m_{\chi^\pm_1}-m_{\chi^0_1}\gsim5\GeV$ and the sneutrino effect is not exceptionally large, then chargino masses up to the kinematical limit are excluded \cite{LEP15}. In our case $m_{\chi^\pm_1}-m_{\chi^0_1}=(28-39)\GeV$ . Therefore we conclude that 
\begin{equation}
m_{\chi^\pm_1}\gsim68\GeV\ .\qquad {\rm (LEP~1.5\ charginos)}
\label{eq:chlimit2}
\end{equation}
Comparing our calculations of the chargino cross section with the plots in Ref.~\cite{LEP15}, we have explicitly verified that this generic result indeed applies for our values of the sneutrino mass. Future runs at LEP~2 energies
of $\sqrt{s}\approx160\,(175)\GeV$ should be able to probe chargino masses up to $80\,(87)\GeV$, which would amount to probing two-thirds (nearly all) of the parameter space. However, signal extraction will be complicated by the significant WW background expected.

Right-handed selectron pair production proceeds via $s$-channel $\gamma,Z$
exchange and $t$-channel neutralino exchange, with the subsequent decay 
$\tilde e_R\to e\chi^0_1$. Our cross sections at LEP~1.5 energies range from 2.5 down to 1.2 pb for the allowed mass range $m_{\tilde e_R}\lsim50\GeV$. At the same time the selectron-neutralino mass difference is in the range $m_{\tilde e_R}-m_{\chi^0_1}=(12-2)\GeV$. Efficiencies for selectron detection are negligible for mass differences below 5 GeV, they are around 55\% for mass
differences of 5--6 GeV, and they climb up to 75\% for 10 GeV or larger mass differences \cite{LEP15}. Taking our calculated cross sections, times the
integrated luminosity ($\approx2.8\ipb$), times these efficiencies, we can
exclude points in parameter space that predict three or more events. We obtain the lower bound $m_{\tilde e_R}>46\GeV$, which can be translated into a (more
useful) lower bound on the chargino mass in this model
\begin{equation}
m_{\chi^\pm_1}\gsim70\GeV\ .\qquad {\rm (LEP~1.5\ selectrons)}
\label{eq:chlimit3}
\end{equation}
Smuon and stau production have smaller cross sections and do not produce
any new constraints on the parameter space. Running at higher energies increases the selectron cross section. With sufficient integrated luminosity one should be able to probe indirectly somewhat higher chargino masses ($m_{\chi^\pm_1}\lsim75\GeV$). To overcome the soft-daugther-lepton problem
when $m_{\tilde e_R}\approx m_{\chi^0_1}$, with sufficient integrated luminosity one could resort to the radiative process $e^+e^-\to\tilde e^+_R\tilde e^-_R\gamma$, employing a hard photon tag \cite{CDG}.

\section{Conclusions}
\label{sec:conclusions}
We have demonstrated the robustness of the experimental predictions of the
model presented in Ref.~\cite{Zero} under changes in various GUT-scale
parameters. We then applied the recent Tevatron and LEP~1.5 data and discovered that one-third of the parameter space of the model has since become excluded,
\ie, $m_{\chi^\pm_1}\gsim70\GeV$ is required. The most immediate prospects for further exploration of the model are to be found in the ongoing analyses of the Tevatron trilepton data and future higher-energy runs at LEP~2. 

\section*{Acknowledgments}
We would like to thank John Ellis, Teruki Kamon, Lee Sawyer, and James White for helpful discussions.
The work of J.~L. has been supported in part by DOE grant DE-FG05-93-ER-40717.
The work of D.V.N. has been supported in part by DOE grant DE-FG05-91-ER-40633.

\section*{Note Added in Proof}
Since the completion of this paper, the CDF Collaboration has performed an
analysis of its trilepton data ($\sim100\ipb$) in the context of the present
model and concluded that $m_{\chi^\pm_1}\gsim70\GeV$ is required \cite{Moriond}. As anticipated, the soft nature of the daughter leptons from
$\chi^0_2$ decay constitute the limiting factor in the chargino mass reach.
Also, the D0 Collaboration has updated its estimate for the top-quark mass
($m_t=170\pm15\pm10\GeV$ \cite{D0top}), which is now in good agreement with
our theoretical prediction in Eq.~(3).

\clearpage

\clearpage

\clearpage

\clearpage

\end{document}